 \documentclass[smallabstract,smallcaptions]{dccpaper}

\usepackage{epsfig}
\usepackage{cite}
\usepackage{amsmath}
\usepackage{amssymb}
\usepackage{color}
\usepackage{url}

\usepackage{amsmath}
\usepackage{graphicx}
\usepackage{subcaption}

\newtheorem{definition}{Definition}
\newtheorem{lemma}{Lemma}

\newlength{\figurewidth}
\newlength{\smallfigurewidth}

\usepackage{tikz}
\usetikzlibrary{shapes}

\tikzset{
  Speaker/.pic={
    \filldraw[fill=gray!40,pic actions] 
    (-15pt,0) -- 
      coordinate[midway] (-front) 
    (15pt,0) -- 
    ++([shift={(-6pt,8pt)}]0pt,0pt) coordinate (aux1) -- 
    ++(-18pt,0) coordinate (aux2) 
    -- cycle 
    (aux1) -- ++(0,6pt) -- coordinate[midway] (-back) ++(-18pt,0) -- (aux2);
  }
  
}

\begin{document}


\title
{\large
\textbf{Rate-Distortion under Neural Tracking of Speech: \\ A Directed Redundancy Approach}
}

\author{%
Jan {\O}stergaard, Sangeeth Geetha Jayaprakash, and Rodrigo Ordoñez\\[0.5em]
{\small\begin{minipage}{\linewidth}\begin{center}
Department of Electronic Systems \\
Aalborg University \\
Fredrik Bajers Vej 7 \\
Aalborg, 9000, Denmark \\
\url{{jo,sgjay,rop}@es.aau.dk} 
\end{center}\end{minipage}}
}

\maketitle
\thispagestyle{empty}

\vspace{1cm}

\begin{abstract}
 The data acquired at different scalp EEG electrodes when human subjects are exposed to speech stimuli are highly redundant. The redundancy is partly due to volume conduction effects and partly due to localized regions of the brain synchronizing their activity in response to the stimuli. In a competing talker scenario, we use a recent measure of directed redundancy to assess the amount of redundant information that is causally conveyed from the attended stimuli to the left temporal region of the brain. We observe that for the attended stimuli, the transfer entropy as well as the directed redundancy is proportional to the correlation between the speech stimuli and the reconstructed signal from the EEG signals. 
 This demonstrates that both the rate as well as the rate-redundancy are inversely proportional to the distortion in neural speech tracking. Thus, a greater rate indicates a greater redundancy between the electrode signals, and a greater correlation between the reconstructed signal and the attended stimuli. A similar relationship is not observed for the distracting stimuli. 
 \end{abstract}

\Section{Introduction}

Biological cognitive systems are capacity limited in the sense 
of having finite memory, processing power, and communication capabilities. Since Shannon's foundational work --- a mathematical theory of communications \cite{shannon:1948}, there has been numerous works addressing the information rate of the human brain from different perspectives. Factors such as synaptic transmission speed, neural network connectivity, and sensory integration of information across different brain regions all play a role in determining the information rate. For an overview of information theory in cognition, see e.g., \cite{e20090706}. 
Whilst the body senses and communicates millions of bits per second of data to the brain, only a fraction of this information is processed by the conscious mind. This data reduction is due to the attentional filter being able to compress information and remove irrelevant information \cite{broadbent:1958}.  
Barlow hypothesized that the brain makes use of efficient coding to compress the information and guarantee that neurons transmit information with minimal redundancy \cite{barlow:1961}. 
The information of spiking neurons were considered in e.g.,\cite{Stevens:1995,Crutchfield:2015,7577753,8278262,8944131}. The information capacity of the human visual system has been addressed in \cite{zhaoping:2006,bates:2019}. The information capacity and the information loss of the human auditory system was considered in \cite{jacobson:1950} and \cite{8579632}, respectively. 
A rate-distortion perspective for the human working memory was considered in \cite{10.7554/eLife.79450,bates:2020} and analytical information theoretic frameworks for neural processing were given in \cite{119739} and \cite{5420278}.   
A partial information decomposition  \cite{Williams:2010} approach was taking in \cite{Gutknecht:2021}.

The field of information rates in the brain is vast and we limit here the scope to works related to neural (cortical) speech tracking. Cortical tracking of speech (CTS) refers to the phenomonen that slow moving cortical oscillations in the delta and theta bands are modulated by the envelope of the attended sound \cite{mesgarani_chang, DING201441}. The auditory regions are particularly synchronized to the oscillations of the envelope of the modulating acoustic stimuli, and the strength of the CTS relates to the phase-locking (or neural tracking) capacity of the auditory cortex \cite{ISSA2024120675}. CTS is used in e.g., auditory attention detection (AAD) problems, where a subject is exposed to multiple simultaneous acoustic stimuli. It is common to have two competing acoustic stimuli such as speech versus speech \cite{sullivan:2015} or speech versus music \cite{simon:2023}. 
The goal is to detect which of the acoustic stimuli the subject is paying attention to based on the subject's cortical activity \cite{mesgarani:2012}. The cortical activity is often represented by non-invasive scalp or in-ear electroencephalogram (EEG) recordings, which are mapped to the envelope-domain of the stimulus using linear or non-linear models \cite{sullivan:2015,10.3389/fnins.2018.00531,alickovic:2019,https://doi.org/10.1111/ejn.13790,Reetzke:2021}. The method of regularized least-squares regression utilizes a linear model that is fitted (trained) to the EEG signal \cite{sullivan:2015}. The resulting models are then correlated with the envelopes of the acoustic stimuli, and the stimulus having the greatest correlation is considered to be the one that the subject is paying attention to.  Information-theoretic limits on the performance of auditory attention decoders were obtained in \cite{10476856}. 

To assess the redundancy in EEG signals, one could potentially use the mutual information measure \cite{cover:2012}. The mutual information is only defined for two sets of signals and is therefore not able to quantify the shared information between three or more signals \cite{yeung:2002}. A lower bound of the redundant information in an arbitrary number of signals were given in \cite{ostergaard:2024a} and a new notion of directed redundancy was recently presented in \cite{ostergaard:2024b}. 

In this paper, subjects were exposed to two competing acoustic stimuli (male and female speech) and asked to focus on one of them (the attended signal) and ignore the other (the distractor). We use linear decoders to reconstruct the stimuli from the signals obtained by scalp electrodes located in the proximity of the left temporal region of the brain. The distortion is quantified by $D=1-|\rho|$, where $\rho$ is the Pearson correlation between the reconstructed signal and the attended stimulus. The amount of redundant information causally conveyed from the acoustic stimuli via the left temporal region and to the reconstructed stimuli is measured using the directed redundancy measure of \cite{ostergaard:2024b}. It is demonstrated that the directed redundancy is inversely proportional to the distortion. Thus, a greater amount of redundant information corresponds to a lower distortion and thereby better neural tracking of the acoustic stimuli. Moreover, an increase in rate from the acoustic stimuli and to the reconstructed stimuli is also inversely proportional to the distortion. 
Interestingly, we do not observe such relationships when assessing the distracting stimuli.

\Section{Notation}
Let $X^n= X_1, X_2, \dotsc, X_n$, where $X_k \in \mathbb{R}$ are random variables. To simplify the notation, we sometimes write $X$ instead of $X^n$. 
We use upper case letters for random variables and lower case letters for their realizations.


\Section{Directed Redundancy}
Let $\mathbb{TE}$ denote Schreiber's Transfer Entropy, which is a directed information measure defined  as \cite{schreiber:2000}:
\begin{align}\label{eq:TE}
\mathbb{TE}(X^n\to Z^n) \triangleq
I(X^{n-1} ; Z_n | Z^{n-1} ). 
\end{align}
From \eqref{eq:TE} it can be noticed that transfer entropy is a variant of conditional mutual information, where the mutual information between the current sample $Y_n$ of the target $Y$  and the past $X^{n-1}$ of the source $X$  is quantified conditioned upon the past $Y^{n-1}$ of the target. 

\begin{figure}[th]
\centering
\begin{subfigure}[b]{0.48\textwidth}
\centering
\includegraphics[width=7cm]{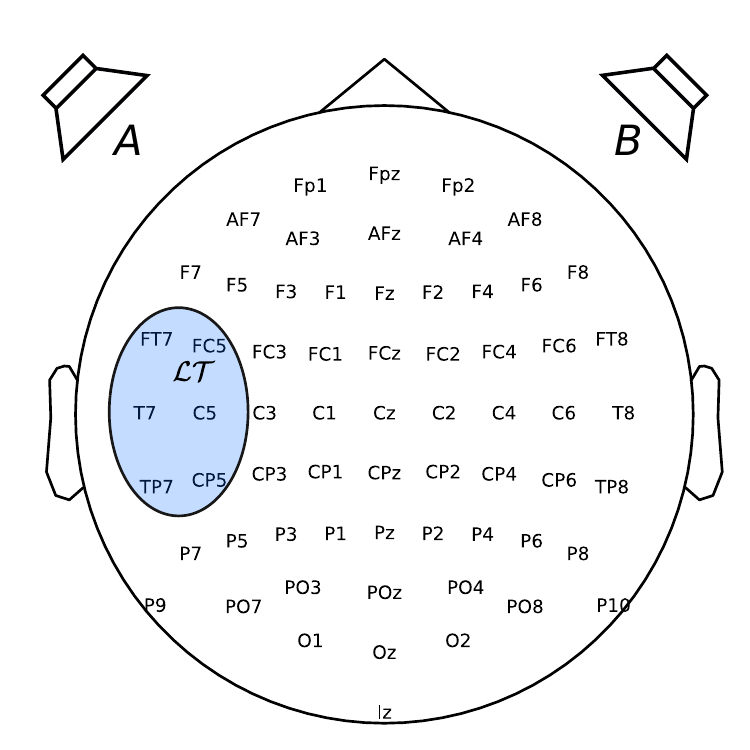}
\caption{A subject is exposed to competing acoustic signals $A$ and $B$ and is focusing on $A$. Labels indicate EEG electrodes. 
The shaded area illustrates the electrodes in $\mathcal{LT}$.}
\label{fig:setup}
\end{subfigure}~%
\hfill
\begin{subfigure}[b]{0.48\textwidth}
        \centering
        \includegraphics[width=7cm]{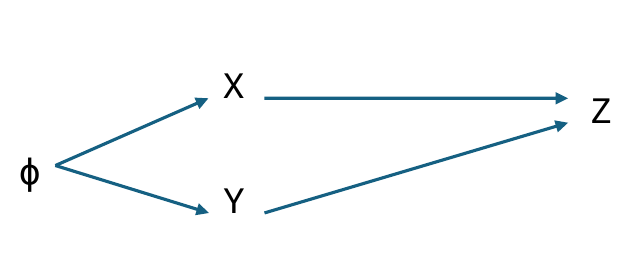}
         \caption{The process $\phi$ drives the redundancy in the system. Information about $\phi$ is causally conveyed to $Z$ via $X$ and $Y$. In our work, $\phi$ models $A$ or $B$,  $X$ and $Y$ represents two of the six electrodes in $\mathcal{LT}$ and $Z$ is the reconstruction of $A$ or $B$ from the electrodes in $\mathcal{LT}$. }
         \label{fig:systam}
\end{subfigure}
\caption{The physical setup considered in this work is illustrated in (a) and a model for the physical setup is shown in (b).}
\end{figure}

Consider the system depicted in Fig.~\ref{fig:systam}. Here $\phi = \phi^n$ is a process that drives the redundancy in the system. We are then interested in quantifying the amount of redundant information about $\phi$ that is causally conveyed via $X$ and $Y$ to $Z$. This is non-trivial to compute since a positive transfer entropy from $X$ to $Z$ and from $Y$ to $Z$ does not mean that $X$ and $Y$ are necessarily sharing the same information with $Z$. Moreover, $X$ and $Y$ are not (necessarily) causally coupled to each other, so we cannot use the transfer entropy from $X$ to $Y$ or from $Y$ to $X$ as a measure of redundant information. In this work, we will instead rely on the causal (directed) redundancy measure of \cite{ostergaard:2024b}, which is formally described by the following definition:

\begin{definition}\cite{ostergaard:2024b}\label{def:red}
\emph{The transfer-entropy measurable causal redundancy provided to the target $Z$ via the source  $X$ and $Y$, which are driven by the hidden redundancy process $\phi$  is defined as:}
\begin{align}\label{eq:min}
    I^{\mathrm{red}}(X;Y;Z) &\triangleq 
    \min \ ( \mathbb{TE}(\phi\to Z), \mathbb{TE}(\phi\to T_{XZ}), \mathbb{TE}(\phi\to T_{YZ}), \\ \notag
      &\qquad \mathbb{TE}(T_{XZ} \to Z),\mathbb{TE}(T_{YZ} \to Z),  I( T_{XZ} ; T_{YZ} ) ), 
\end{align}
\emph{where $T_{XZ}$ is a  \emph{causal} minimal sufficient statistics for $X$ with respect to $Z$ so that $T_{XZ}$ satisfies the conditional Markov chain}
\begin{equation*}
\vspace{-3mm}
X^{n-1}|_{Z^{n-1}} - T_{XZ}^{n-1}|_{Z^{n-1}} - Z_n|_{Z^{n-1}}, \forall n,
\end{equation*}
where
\begin{align*}
T^{n-1}_{XZ} &= \arg\min_{S_{XZ}^{n-1}} I(S_{XZ}^{n-1}; Z_n | Z^{n-1}) \\
& s.t.\quad I(S_{XZ}^{n-1}; Z_n | Z^{n-1}) = I(X^{n-1} ; Z_n | Z^{n-1}), \forall n. 
\end{align*}
\end{definition}

In Definition~\ref{def:red}, the causal sufficient statistics of $X$ and $Y$ with respect to $Z$ are included to make sure that only information that is relevant to $Z$ is taken into account. 
One can allow additional input to $X$ and $Y$, e.g., via feedback from $Z$ or other sources directly interacting with $X$ and $Y$. In such a case, one can further include the terms $I(\phi ; T_{XZ})$ and $I(\phi ; T_{YZ})$ within the minimization of \eqref{eq:min} to make sure only information relevant to $\phi$ is taken into account. 

The causal sufficient statistics, $T_{XZ}$ and $T_{YZ}$ are generally hard to compute and it was suggested in \cite{ostergaard:2024b} to replace them by their corresponding processes. This would then lead to an upper bound on the directed information. 
\begin{lemma}\cite{ostergaard:2024b}\label{lem:ub1}
\emph{Consider the dynamical system in Fig.~\ref{fig:systam}. The amount of redundant transfer entropy communicated via $X$ and $Y$ to $Z$, and contained in $Z$, is upper bounded by:}
\begin{align}\notag
I^{\mathrm{red}}(X& ;Y; Z)  \leq \min\, ( \mathbb{TE}(\phi\to Z), \mathbb{TE}(\phi\to X), \\ 
\label{eq:lem1}
&\mathbb{TE}(\phi\to Y), \mathbb{TE}(X\to Z),\mathbb{TE}(Y\to Z)).
\end{align}
\end{lemma}

In the above example, we considered two processes $X$ and $Y$. In general, we could have $K$ processes $X(1), \dotsc, X(K)$, where $X(i) = X_1(i),X_2(1), \dotsc, X_n(1)$. 

\Section{Auditory Attention Decoding}
At any given time, subjects are presented with two competing acoustic signals $A$ and $B$ as illustrated in Fig.~\ref{fig:setup}. 
These signals are represented as time series using the convention $A^n = A_1,A_2, \dotsc, A_n$, where $A_i \in \mathbb{R}$ for $i=1,\dotsc, n$. A similar notation applies to $B$. We will only be using the envelopes of the acoustic signals, which we will be denoting $S=S^n$. It will be clear from context whether $S$ is the attended signal or the distractor.  

 While the subject is listening to the stimuli, the brain response is recorded via EEG electrodes that are spatially distributed on the scalp. 
We will be focusing on the left-temporal ($\mathcal{LT}$) region, which cover the left auditory cortex and have been shown to be good for cortical tracking of speech \cite{Issa:2024}. We define $\mathcal{LT}$ as the following set of electrodes in the extended 10-20 international 64 channel EEG electrode layout convention \cite{Niedermeyer:2004}:
\begin{align}
    \mathcal{LT} &= \{\mathrm{FT}_7, \mathrm{T}_7, \mathrm{TP}_7, \mathrm{CP}_5, \mathrm{FC}_5, \mathrm{C}_5 \}.
\end{align}

\SubSection{Stimulus reconstruction}
Let $s$ denote the envelope of the stimulus $A$. To reconstruct the stimulus $s$ from the neural responses, we use a backward stimulus-response mapping that exploits all neural responses in a given time frame \cite{sullivan:2015}. Specifically, let $\hat{s}_t$ be the reconstructed stimuli at time $t$ and let $r(t,c)$ denote the instantaneous neural response sampled at time $t$ at EEG electrode $c$. Then, if using a linear decoder $g(\tau,c)$, the reconstructed stimuli can be written on the following form \cite{crosse2016mtrf}:
\begin{equation}\label{eq:recon}
    \hat{s}_t = \sum_{c}\sum_{\tau} r(t+\tau,c) g(\tau,c),
\end{equation}
where the outer sum runs over the number of EEG electrodes and the inner sum over time delays $\tau$ in a desired range of delays to accommodate the non-instantaneous effects of the acoustic stimuli. The optimal linear  decoder $\mathbf{g}$ that minimizes the mean square error: $\sum_t (s_t - \hat{s}_t)^2$ can be computed by \cite{crosse2016mtrf}:
\begin{align}
    \mathbf{g} = (R^TR + \lambda I)^{-1}R^Ts,
\end{align}
where $\lambda$ is a regularization term, $R$ is a matrix containing the lagged time series of the response $r(t+\tau,c)$, and $\mathbf{g}$ is the vector containing the optimal decoder coefficients \cite{crosse2016mtrf}.

In the sequel, we denote by $\hat{s}=\hat{s}_1,\dotsc, \dotsc, \hat{s}_n$ the reconstructed stimuli obtained using the 6 electrodes in $\mathcal{LT}$. We assume that $s$ and $\hat{s}$ are normalized to zero-mean and unit-variance, which means that the stimulus-reconstruction distortion in terms of the mean squared error is an affine function of  the correlation between the signals, that is:
\begin{align}
    \frac{1}{2}\mathbb{E}[\|\hat{S} - S\|^2] = 1- \rho,
\end{align}
where $\rho = \mathbb{E}[\hat{S}S]$. To avoid that negative correlations penalizes the distortion, we  use the following distortion measure:
\begin{equation}\label{eq:distortion}
D\triangleq 1-|\rho|.
\end{equation} 

\Section{Rate Redundancy in EEG Signals}

Let $E^n(j) = E_1(j), E_2(j),\dotsc, E_n(j)$, $j\in \mathcal{LT}$, denote the time-series representation of the $j$-th EEG electrode signal. 

The minimum rate from the electrodes in $\mathcal{LT}$ to the reconstructed stimuli is then found as:
\begin{align}\label{eq:Retohs}
    R_{E\to \hat{S}} \triangleq \min_{j\in \mathcal{LT}} \mathbb{TE}( E^n(j) \to \hat{S}^n).
\end{align}
We note that $\hat{S}^n$ is a deterministic function of $(E^n(1), \dotsc, E^n(|\mathcal{LT}|))$ according to \eqref{eq:recon}, which implies that $I(E^n(1), \dotsc, E^n(|\mathcal{LT}|) ; \hat{S}^n) = 0$ or $I(E^n(1), \dotsc, E^n(|\mathcal{LT}|) ; \hat{S}^n) = \infty$. This is not a problem in this work, since $\hat{S}^n$ is a non-deterministic function of the individual electrode signals, which implies that $I(E^n(j) ; \hat{S}^n)$ is bounded for all $j=1,\dotsc, |\mathcal{LT}|$.

The minimum rates from the acoustic stimuli and to the electrodes are given by:
\begin{align}\label{eq:Rse}
     R_{S\to E} \triangleq \min_{j\in \mathcal{LT}} \mathbb{TE}(
     S^n \to E^n(j) ).
\end{align}
Finally, the rate from the stimulus to the reconstructed stimulus is given by:
\begin{align}\label{eq:Rss}
    R_{S\to \hat{S}} \triangleq \mathbb{TE}(
     S^n \to \hat{S}^n ).
\end{align}

The directed redundancy for the brain region is then upper bounded by $R$, which is found using Lemma~\ref{lem:ub1}, that is:
\begin{align}\label{eq:RminLT}
   R \triangleq\min ( R_{S\to \hat{S}},     R_{E\to \hat{S}} , R_{S\to E} ).
\end{align}

\Section{Simulation Study}
We use the EEG signals described in detail in \cite{fuglsang:2017}. 
The subjects were exposed to competing acoustic stimuli in the form of male and female talkers. 15 subjects each having 60 repeated trials were used. The temporal envelopes of the target and distractor were acquired and downsampled to 64 Hz to match the sampling frequency of the EEG signals.

In order to do stimulus reconstruction, we trained subject dependent linear decoders using EEG data only from the electrodes in $\mathcal{LT}$. The decoders for reconstructing the stimuli were obtained using the mTRF-Toolbox~\cite{crosse2016mtrf}. Separate decoders were trained for the attended and distracting stimuli.

\begin{figure}
    \centering
        \begin{subfigure}[t]{0.5\textwidth}
        \centering
    \includegraphics[width=6cm]{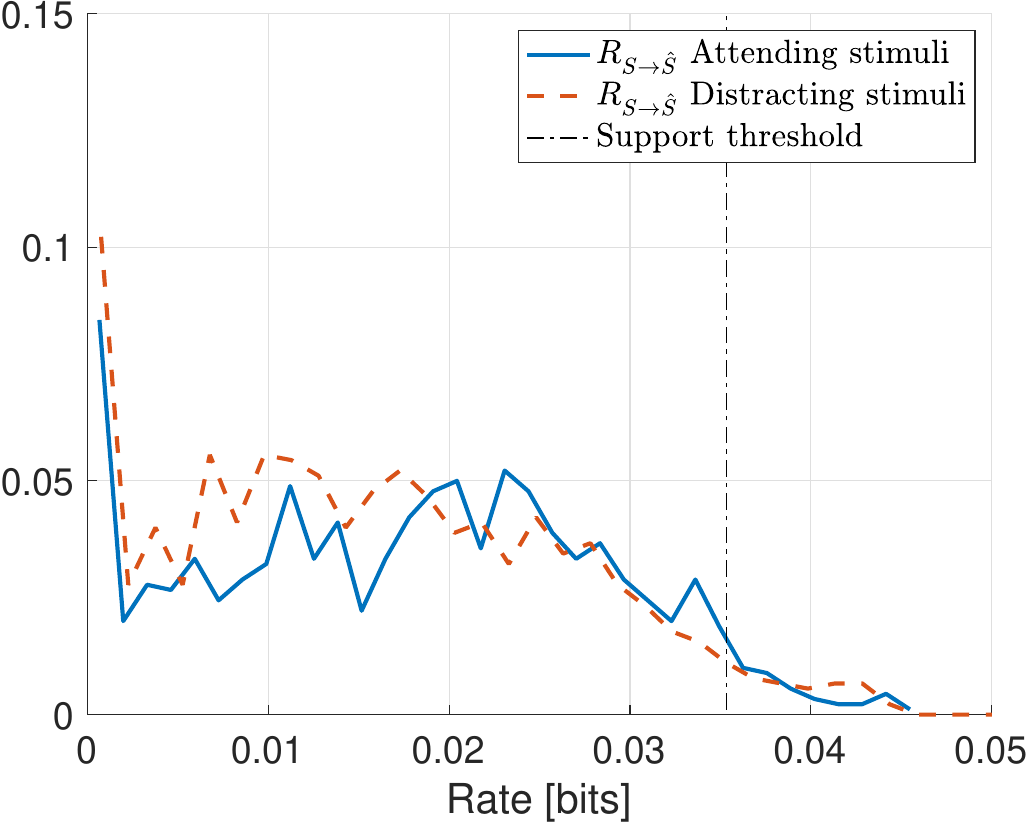}
    \caption{$R_{S\to\hat{S}}$ \eqref{eq:Rss}}
    \label{fig:pdf_R}
     \end{subfigure}%
    ~ 
    \begin{subfigure}[t]{0.5\textwidth}
        \centering
   \includegraphics[width=6cm]{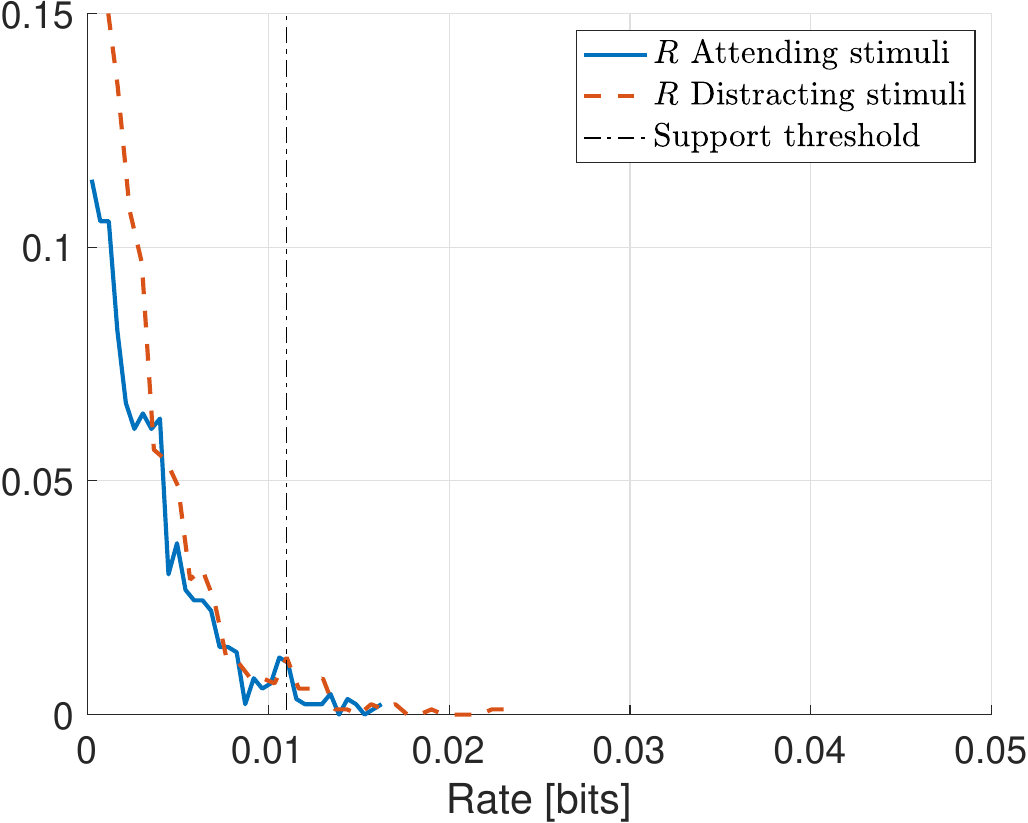}
    \caption{$R$ \eqref{eq:RminLT}}
    \label{fig:pdf_R1}
     \end{subfigure}%
     \\
    \begin{subfigure}[t]{0.5\textwidth}
        \centering
    \includegraphics[width=6cm]{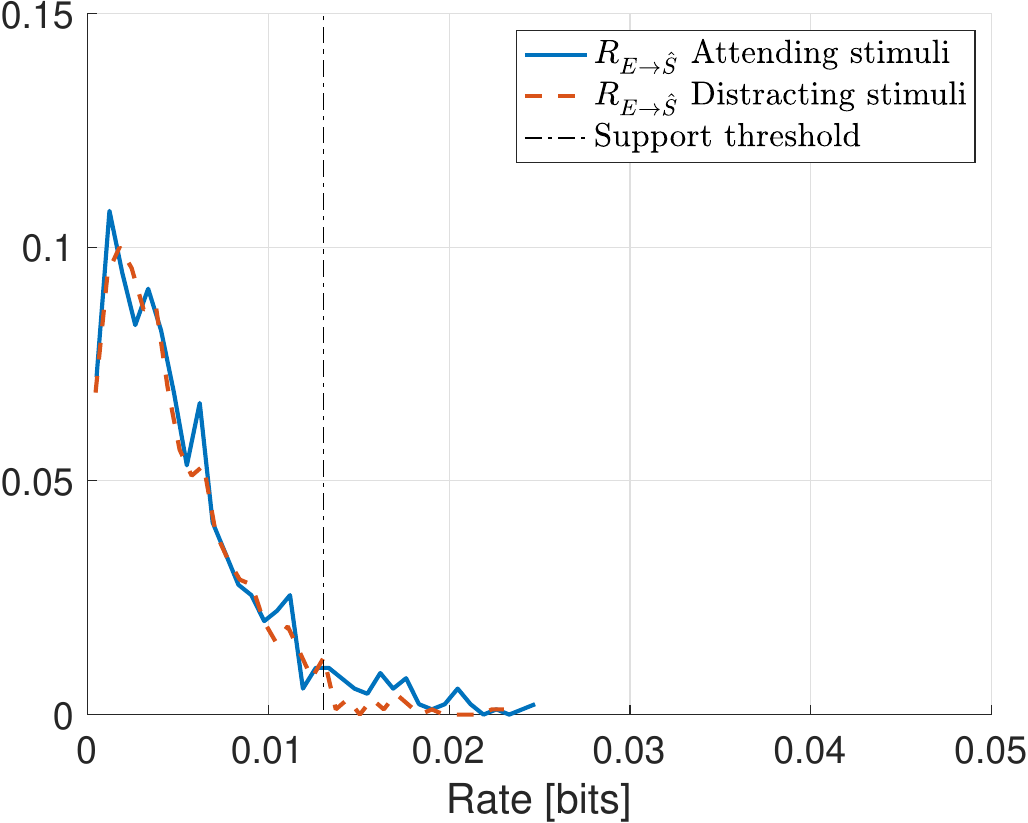}
    \caption{$R_{R\to \hat{S}}$ \eqref{eq:Retohs}}
    \label{fig:pdf_R2}
     \end{subfigure}%
    ~ 
    \begin{subfigure}[t]{0.5\textwidth}
        \centering
   \includegraphics[width=6cm]{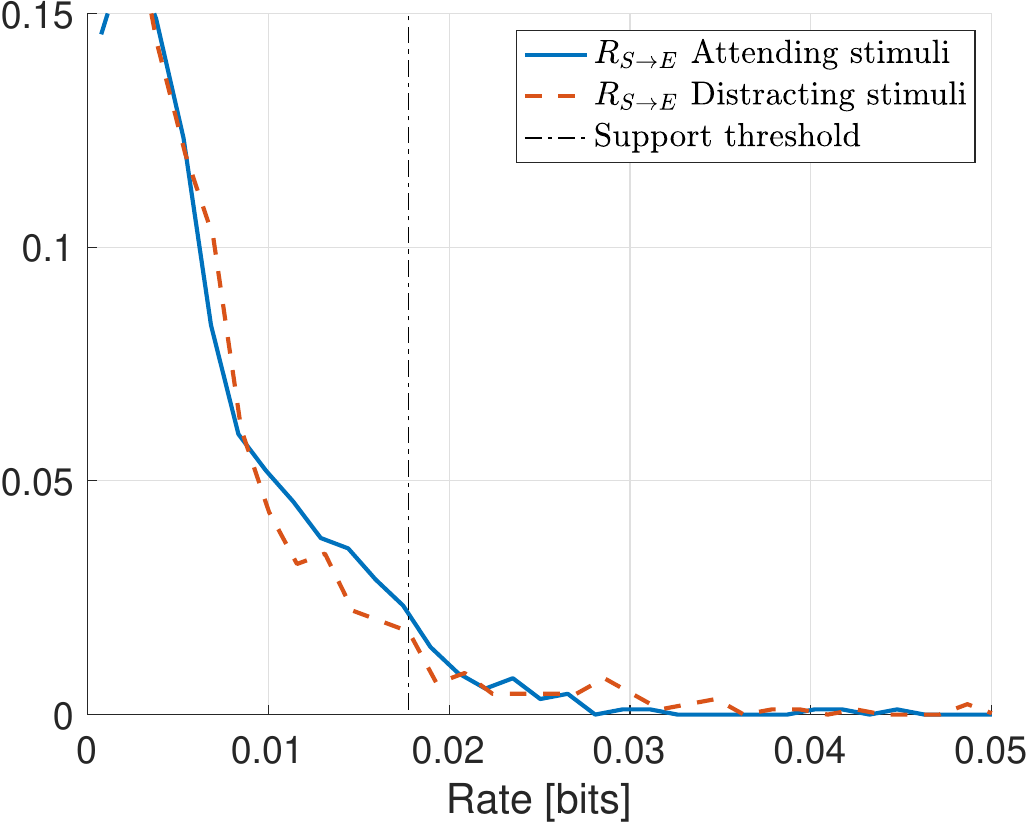}
    \caption{$R_{S\to E}$ \eqref{eq:Rse}}
    \label{fig:pdf_R3}
     \end{subfigure}%
     \caption{Probability density functions of the rates \eqref{eq:Retohs}, \eqref{eq:Rse}, and \eqref{eq:Rss} as well as for the directed rate redundancy \eqref{eq:RminLT}. The two cases of attending stimuli (blue) and distracting stimuli (dashed red) are shown.}
\end{figure}

The probability density functions of the rates \eqref{eq:Retohs}, \eqref{eq:Rse}, and \eqref{eq:Rss} as well as for the directed rate redundancy \eqref{eq:RminLT} are shown in Figs.~\ref{fig:pdf_R} -- \ref{fig:pdf_R3}. Due to the minimization operation invoked in \eqref{eq:Retohs}, \eqref{eq:Rse}, and \eqref{eq:RminLT}, the support of the these informational rates are less than for \eqref{eq:Rss}. In general, greater rates are increasingly unlikely to appear. In these figures, the vertical dashed-dotted black lines indicate the maximum rate at which the corresponding pdf is above $0.01$. In the sequel, we will only be using distortion points, where the associated rates are within the \emph{support} region, i.e., on the left side of this support threshold. We note that rate points outside this support region have very few associated distortion points, and the distortion-rate estimate for these rates would therefore be very poor.

\begin{figure}[t!]
    \centering
    \begin{subfigure}[t]{0.5\textwidth}
        \centering
        \includegraphics[width=7cm]{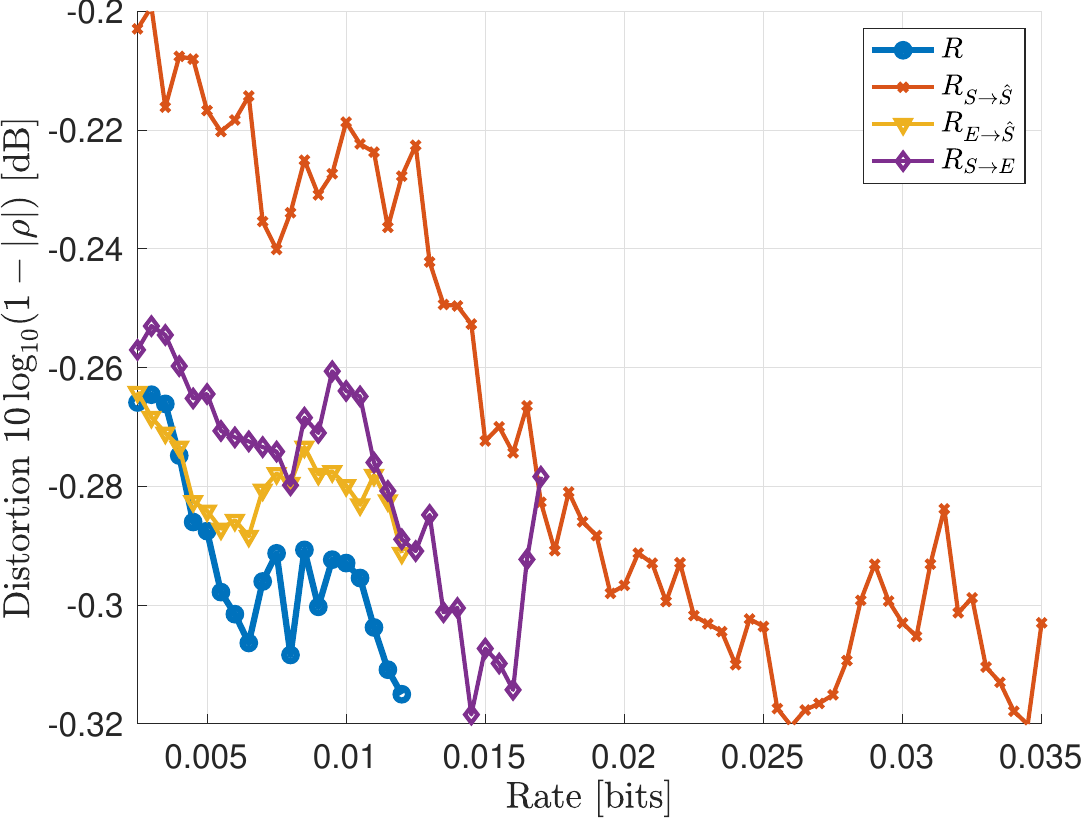}
        \caption{Attending stimuli}
        \label{fig:rd-att}
    \end{subfigure}%
    ~ 
    \begin{subfigure}[t]{0.5\textwidth}
        \centering
        \includegraphics[width=7cm]{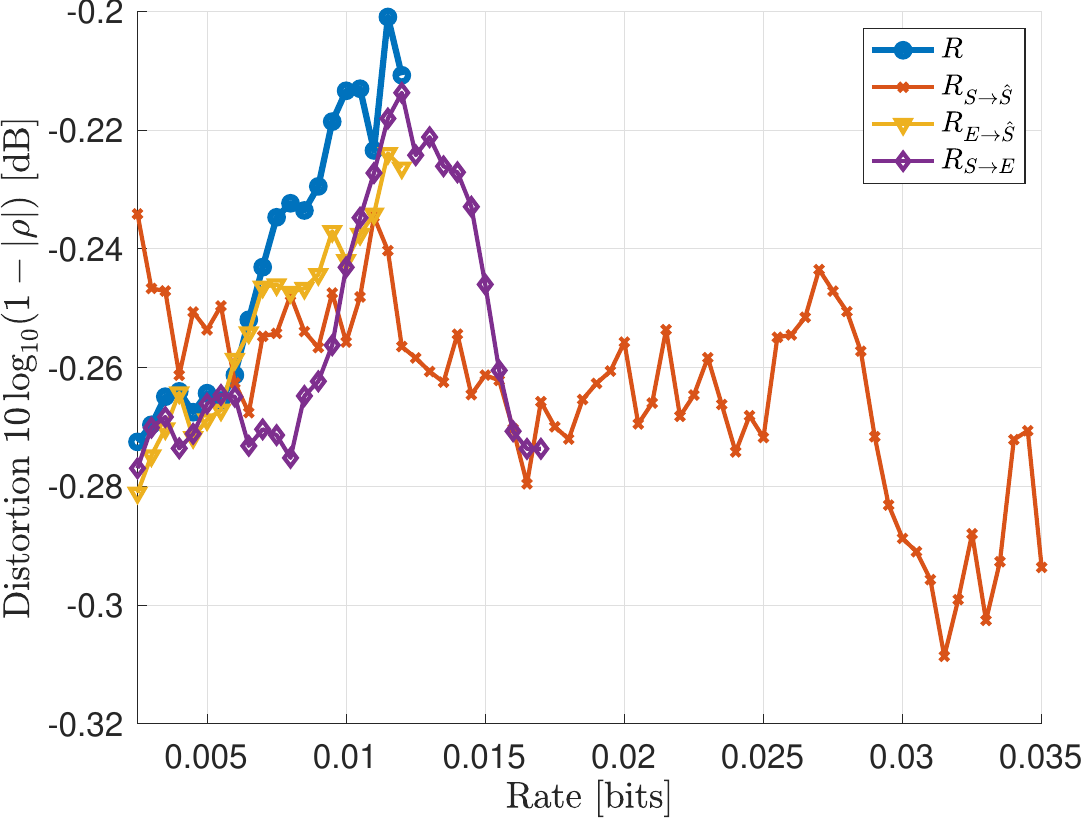}
        \caption{Distracting stimuli}
        \label{fig:rd-dst}
    \end{subfigure}
    \caption{Operational distortion-rate curves as a function of rates \eqref{eq:Retohs}, \eqref{eq:Rse}, and \eqref{eq:Rss} as well as the directed redundancy \eqref{eq:RminLT}.}
    \label{fig:all}
\end{figure}

In Figs.~\ref{fig:rd-att} and \ref{fig:rd-dst}, we have shown the distortion \eqref{eq:distortion} (in dB) as a function of the different types of rates considered. Specifically, we have shown the distortion as a function of $ R_{E\to S}$ \eqref{eq:Retohs}, $R_{S\to E}$ \eqref{eq:Rse}, and $R_{S\to \hat{S}}$ \eqref{eq:Rss}  as well as $R$ \eqref{eq:RminLT}. This is shown for the attended stimuli in Fig.~\ref{fig:rd-att} and for the distracting stimuli in Fig.~\ref{fig:rd-dst}. 
The distortion-rate curves are obtained by splitting the rate into overlapping intervals of 0.005 bits, and averaging all the distortion points that are associated with rates within the same bin. Only rates within the support regions are considered.

From Fig.~\ref{fig:rd-att}, it can be observed that the distortion shows a decreasing trend as a function of the 
transfer entropy $R_{S\to\hat{S}}$ from the attended stimuli to the reconstructed signal. Such a trend is less clear for the distracting stimuli in Fig.~\ref{fig:rd-dst}. 
From the figures, it can also be seen that the distortion shows a decreasing (increasing) trend as a function of the directed redundancy $R$ for attended (distracting) stimuli. 
Thus, for the attended stimuli, we generally observe that greater correlations (smaller distortions) are approximately proportional to the amount of transfer entropy measured between the acoustic signal and the reconstructed signal. Moreover, there is also an increase in the redundant information $R$ about $S$, which is conveyed by the electrode signals to $\hat{S}$. Thus, a greater information flow from the attended stimuli to the reconstructed stimuli also leads to greater redundancy in the EEG signals.  On the other hand, for the distracting stimuli, the directed redundancy appears to be inversely proportional to the distortion. Thus, an increase in the speech tracking performance of the distracting stimuli corresponds to a decrease in the amount of redundant information shared by the EEG electrodes. 

To assess the strength of the potentially linear relationship between the distortion (expressed in dB) and the rates and directed redundancy, we have fitted a linear regression model to the corresponding rate-distortion points. We use the fitlm command in Matlab, and test the null hypothesis that the model parameters of the models are zero. The corresponding $p$-values are reported in Table \ref{tab:p-values}. The boldfaced p-values indicate the cases for which a linear model of the rates is a statistically significant (less than 0.05) predictor for the corresponding distortion. 
 It can be seen that  $R$  is a statistical significant linear predictor for the distortion both in case of attended or distracting stimuli.
 Moreover, in the case the attending stimuli, the slopes are negative, and for the distracting stimuli, the slope is positive, cf.\ Figs.~\ref{fig:rd-att} and ~\ref{fig:rd-dst}.
 
\begin{table}[th]
    \centering
    \caption{$p$-values  when fitting a linear model to the rate-distortion points. Bold faced values are significant ($p<0.05$).}
    \begin{tabular}{|c|p{2cm}|p{2.2cm}|p{2.2cm}|p{2.2cm}|} \hline
                    &   $R_{S\to\hat{S}}$ & $R_{S\to E}$ & $R_{E\to S}$ & $R$ \\ \hline
        Attended & \textbf{6.1e-06}  &  0.0930 & 0.2387& \textbf{0.0177} \\ \hline 
        Distractor & 0.1886 &  \textbf{0.0448} & 0.0553 & \textbf{0.0416} \\ \hline 
    \end{tabular}
    \label{tab:p-values}
\end{table}

\Section{Conclusions}
We demonstrated using real-world EEG data that the distortion in an auditory attention decoding study is linearly related to the rate (transfer entropy) from the acoustic stimuli to reconstructed stimuli using the brain response from the left temporal region. Moreover, the distortion is also linearly related to the directed redundancy of the EEG electrode signals in the left temporal region. This suggests that a greater correlation between the envelope of the speech stimuli and the reconstructed stimuli is obtained when a greater amount of redundant information about the stimuli is conveyed in the brain signals.

\Section{References}


\end{document}